\documentclass{aastex}          
\usepackage{spr-astr-addons}    

\begin{document}
%
\title{Intermediate evolution using SNIa, and BAO}

\shorttitle{Intermediate evolution}

\shortauthors{V.H.Cardenas and O.Herrera}

\author{V\'ictor H. C\'ardenas and O. Herrera }
\affil{Instituto de F\'{\i}sica y Astronom\'ia, Facultad de
Ciencias, Universidad de Valpara\'iso, Av. Gran Breta\~na 1111,
Valpara\'iso, Chile}


\begin{abstract}
We study the intermediate evolution model and show that, compared
with the recent study of a power-law evolution, the intermediate
evolution is a better description of the low-redshift regime
supported by observations from type Ia supernovae and BAO. We found
also that recent data suggest that the intermediate evolution is as
good a fit to this redshift range as the $\Lambda$CDM model.
\end{abstract}

\keywords{cosmology; dark energy}

\section{Introduction}

The current cosmological paradigm was set primarily after
discovering that type Ia supernovae (SNIa), a well-studied standard
candle, are dimmer than expected in the context of the standard
cosmological model \citep{snia1,snia2}. The simplest way to describe
this observation in the context of the standard big-bang picture
is by introducing a cosmological constant term, $\Lambda$, into
Einstein's field equations.

In this setup the cosmological constant $\Lambda$ drives the current
accelerated expansion of the universe, leading to the successful
$\Lambda$CDM model, the simplest one that fits a varied set of
observational data such as SNIa, measurements of the baryon acoustic
oscillations (BAO), information from the cosmic microwave background
radiation (CMBR), growth of structure, etc \citep{weinb12, pero}.

However, this model looks unnatural in at least two ways; (i) there
is no clue about the physical mechanism to get the current value for
$\Omega_{\Lambda}$, and (ii) this model tells us we live in a very
special epoch where the cosmological constant contribution
$\Omega_{\Lambda}$ is of the same order of magnitude as the dark
matter contribution $\Omega_M$, a highly improbable fact
considering the dark matter contribution decreases as $a^{-3}$, with
$a(t)$ the scale factor. Meanwhile the cosmological constant
contribution is and always has been a constant.

A plausible alternative is to explore deviations from the
$\Lambda$CDM model, assuming a variable $\Lambda$. Dark energy (DE)
is the name of the unknown component responsible for the current
accelerated expansion of the universe \citep{dereview}. In its
simplest form, this can be described by a fluid with constant
equation of state (EoS) parameter $w=-1$ corresponding to a
cosmological constant. There are also models where a scalar field
drives the cosmic acceleration, so-called quintessence models
\citep{quinta}, as well as models where through a modification of the
gravity sector the cosmic acceleration is described \citep{modgrav}.

This honest approach focuses only on problem (ii), mentioned in the
previous paragraph, but it is almost completely disconnected from
problem (i). Most of the work in cosmology since the discovery of
the accelerated expansion of the universe has been of this type.
The efforts are focused on characterizing this new component, dark
energy (DE), somehow disconnected from first principles.

Many scientists think that a real understanding of the DE problem,
focusing this time on problem (i), i.e., the old well-known
cosmological constant (C.C.) problem \citep{cc1,cc2}, would
certainly reveal a door to new physics.

Among the many attempts to tackle the cosmological constant problem,
one of the most interesting is where a suitable relaxation mechanism
drives the C.C. to a small value. For example, \cite{Dolgov:1996zg}
describes a model where a scalar field is coupled to the curvature
of the space-time, such that their contribution to the energy
density cancels the vacuum energy. Similar to this work are those
proposed by \citet{Hebecker:2000au}, where an additional scalar
field is introduced, which to date remains a C.C. which
asymptotically vanishes (see \citet{Padmanabhan:2002ji} for a review).

In a recent paper \citep{Dolgov:2014faa} the authors used a
power-law evolution $a(t) \simeq t^{\beta}$ -- suggested by modified
gravity theories at low redshift -- and directly studied the
constraints the supernova data imposed on the parameter $\beta$.
They found, surprisingly, using both data from the Union 2.1
\citep{suzuki} and the recently released joint light-curve analysis
(JLA) set \citep{jla}, that SNIa data suggest a $\beta \simeq 3/2$
assuming a flat universe. Although successful in producing a good
fit to the SNIa data, the model is unable to describe a transition
from a decelerated to an accelerated phase. In fact, by using a
scale factor $a(t) \simeq t^{\beta}$, this implies a constant
deceleration parameter $q=(1-\beta)/\beta$. This transition is a key
ingredient that any model must satisfy if it claims to describe
the recent (low redshift $z<1$) evolution of the universe, which is
essentially the same redshift range as the supernova data span.

It is the purpose of this paper to explore a slightly different
scale factor evolution -- the so-called intermediate evolution --
that interpolates between a power-law evolution and an exponential
one, that is able to describe the transition from the decelerated
phase to an accelerated one. We use the latest data sets of SNIa
(some of which was used in \citep{Dolgov:2014faa}), and compare our
results.

In the next section we introduce the intermediate evolution. In
section III we describe the results of our study and the comparison
with previous works. We end with a discussion section.

\section{The intermediate evolution}

The intermediate evolution was introduced first in the context of
inflationary models as an exact solution for a scalar field
potential of the type $ V(\phi) \propto \phi^{-4(f^{-1}-1)}$
\citep{var05,var051,var052}, where $f$ is a free parameter with
range $0 < f < 1$. A potential of this form, in the context of
the slow-roll approximation, gives a Harrison-Zel'dovich spectrum of
density perturbations with an exact scale-invariant spectral index
i.e., $n_s = 1$.

The motivation to study an intermediate inflationary model comes from
string/M theory. This theory suggests that in order to have a
ghost-free action, high order curvature invariant corrections to the
Einstein-Hilbert action must be proportional to the Gauss-Bonnet
(GB) term \citep{BS85,BS851}. GB terms arise naturally as the
leading order of the expansion to the low-energy string effective
action \citep{KM07,KM071}. This kind of theory has been applied to a
possible resolution of the initial singularity problem
\citep{ART94}, to the study of Black-Hole solutions
\citep{var06,var061,var062}, and accelerated cosmological solutions
\citep{var07,var071,var072,BLP06,delCampo:2014toa}, among others.

Particularly interesting to this work is the finding that, using a
GB interaction with a scalar field $\phi$, it is possible to
describe a DE model leading to a solution of the form $a(t) = a_0
\exp{\left[\left(2/(\kappa\,n)\right)\,t^{1/2}\right]}$
\citep{S07,S071}. Here, $\kappa = 8\,\pi\,G$ and $n$ is an arbitrary
constant. Actually, this is exactly a particular case of the
intermediate evolution, in which the scale factor evolves as
\begin{equation}\label{a}
a(t) = b \exp{\left(At^f\right)},
\end{equation}
where $A$ is a positive constant and $f$ was introduced above. Thus, the expansion of the universe is slower than standard
de Sitter inflation ($a(t) = \exp{(Ht)}$), but faster than power law
inflation ($a(t) = t^p; p > 1$).

Using (\ref{a}) the Hubble function is
\begin{equation}\label{h}
H=\frac{\dot{a}}{a}=Af t^{f-1},
\end{equation}
and the deceleration parameter is
\begin{equation}\label{q}
q=-\frac{\ddot{a}a}{\dot{a}^2} = -1 + \frac{1-f}{f} \frac{1}{A t^f}.
\end{equation}
This last result shows immediately that this kind of evolution
makes it possible to describe the transition from a decelerated to an
accelerated expansion phase. In fact, for small $t$ the second term
dominates, which for $f<1$ gives a decelerating universe evolution,
and for late (larger) times, the first factor starts to dominate,
leading to an accelerated evolution.

From a purely theoretical point of view, most of the efforts made
using the GB scenario to tackle the DE problem, have focused on
obtaining solutions which look similar to the $\Lambda$CDM model at
the background level (see  \citep{Granda:2014zea} and references
therein). A very interesting result was found in
\citep{DeFelice:2009rw}, where by studying the so-called general GB
gravity, the authors show in the linear perturbations regime that
there is an instability during the radiation and matter domination
epoch. This modified GB gravity is equivalent to the term GB
coupled to a scalar field, but without a kinetic term. There remains
the uncertainty as to whether this instability appears once we turn on the
kinetic term, as is the case of the model from which the
intermediate evolution emerged. Although this instability does not
spoil out the power of the model to describe DE, as the authors
mentioned in the text, it is clear that this issue must be delved into more deeply and also in a more general setup (see for
example \citet{delaCruzDombriz:2011wn}).

More interesting for the present work are the results from a direct
test against observational data. In \citep{KM07} the authors use
data from solar system, type Ia supernova, cosmic background
radiation, large-scale structure and nucleosynthesis. They found
some tension with nucleosynthesis and the baryon acoustic scale.
However, their results are based on working in the special case of
an exponential potential $V(\phi)$ for the scalar field. The
intermediate evolution exists for a more elaborate scalar field
potential, and this fact makes a study of this
type of evolution imperative.

To test the model against the observation, it is useful to write
down all the formulae in terms of the redshift, $z=-1 +
a(t_0)/a(t)$. We also impose that $a(t_0)=1$, given the relation
\begin{equation}\label{cons1}
\ln b(1+z)= -At^f.
\end{equation}
From this equation we can rewrite the deceleration parameter
(\ref{q}) as
\begin{equation}\label{qdz}
q(z) = -1 + \frac{f-1}{f} \frac{1}{\ln b (1+z)},
\end{equation}
and the Hubble function can be written as
\begin{equation}\label{hdz}
H(z)=H_0 \left[ 1 + \frac{\ln(1+z)}{\ln b} \right]^{\frac{f-1}{f}}.
\end{equation}
Usually the ratio $E(z)=H(z)/H_0$ is what we need to perform the
statistical analysis. It is clear that the free parameters of the
model are: $h$, $f$ and $b$. The original parameter $A$ in (\ref{a})
is related to $b$ through the age of the universe $t_0$, by $\ln b =
-A t_0^f$.

Because this model shows a transition from a decelerated phase to an
accelerated one, we choose to replace the $b$ parameter by the cross
redshift $z_c$, at which $q=0$. From (\ref{qdz}) we get
\begin{equation}\label{zc}
\ln b = \frac{f-1}{f} - \ln (1+z_c).
\end{equation}
\

In the next section we shall use (\ref{hdz}), written in terms of
the free parameters $h$, $f$ and $z_c$, to test the model against
SNIa data, using both the Lick Observatory Supernova Search (LOSS)
compilation set \citep{Ganeshalingam:2013mia} comprised of $586$
SNIa, and the Joint Ligh-curve Analysis (JLA) set \citep{jla} with
740 points, the largest compilation so far.

We also add baryon acoustic oscillation (BAO) data points from
\citep{baos}, using the approach considered by \cite{xia12} and
\cite{Dolgov:2014faa}. From the quoted references it is found that:
\begin{equation}\label{dvs}
d(z) = (0.335 \pm 0.016, 0.576\pm 0.022, 1.539\pm0.039),
\end{equation}
for $z=(0.106,0.2,0.57)$ respectively, where $d(z)\equiv
D_V(z)/D_V(z=0.35)$ with
\begin{equation}\label{dvdef}
D_V(z)=\left[(1+z)^2 D_A^2(z) \frac{cz}{H(z)}\right]^{1/3},
\end{equation}
where $D_A(z)$ is the angular diameter distance defined by
\begin{equation}\label{da}
D_A(z)=\frac{c}{H_0(1+z)}\int_0^{z} \frac{dz'}{E(z')},
\end{equation}
valid for flat space.

\section{Testing the model}

In this section we use the Loss compilation set
\citep{Ganeshalingam:2013mia} and the JLA set \citep{jla}, together
with the BAO points to constrain the free parameters for the intermediate
evolution model.

In the case of the Loss compiled set \citep{Ganeshalingam:2013mia},
the fitting is done by minimizing the $\chi^2$ function,
\begin{equation}\label{chi2loss}
\chi^2 = \sum_{i} \frac{(\mu_i - \mu_{th}(z_i))^2}{\sigma_i^2},
\end{equation}
where $\mu_i$, $\sigma_i$ are the observational values of the
distance modulus and its errors, while $\mu_{th}(z_i)$ is the
theoretical value of the distance modulus evaluated at the observed
redshift $z_i$,
\begin{equation}\label{musn}
\mu_{th}(z)=m-M=5 \log_{10}D_L(z) + 25,
\end{equation}
where
\begin{equation}\label{dl}
D_L(z)=\frac{c}{H_0}(1+z)\int_0^{z} \frac{dz'}{E(z')},
\end{equation}
is the luminosity distance in flat space.

In the case of the JLA set \citep{jla}, the function we minimize is
\begin{equation}\label{chi2jla}
\chi^2 = (\mu - \mu_{th})^{T} C^{-1}(\mu - \mu_{th}),
\end{equation}
where $C$ corresponds to the covariance matrix released in
\citep{jla}, and the distance modulus is assumed to take the form
\begin{equation}\label{mujla}
\mu = m - M + \alpha X - \gamma Y,
\end{equation}
where $m$ is the maximum apparent magnitude in the rest frame of the
B band, $X$ is related to the time stretching of the light-curves,
and $Y$ corrects the color at maximum brightness. In the general
case, cosmology is restricted together with the parameters $ M $,
$X$ and $ Y $. In \citep{jla} is also given a compressed form of the
data, where only $ M $ is left as a free parameter.


\subsection{$\Lambda$CDM model}

In order to have a reference model to compare with, we start
studying the performance of the $\Lambda$CDM model against the JLA
SNIa data set \citep{jla}. We assume a $\Lambda$CDM model with
arbitrary curvature to ensure three free parameters in the fitting
process; $h$, $\Omega_m$ and $\Omega_{\Lambda}$. The Hubble
parameter for this model is
\begin{equation}\label{hlcdm}
E^2(z)=\Omega_m (1+z)^2z-\Omega_{\Lambda}z(2+z)+(1+z)^2.
\end{equation}
The best fit gives $\chi^2_{min}=32.76$. After marginalizing in $h$,
the following best fit parameters are obtained: $\Omega_m=0.19\pm
0.11$, $\Omega_{\Lambda}=0.56\pm 0.17$, which agree with the values
reported in \citep{jla} and in particular with Figure 15 in that paper.

\begin{figure}[h!]
\begin{center}
\includegraphics[width=5cm]{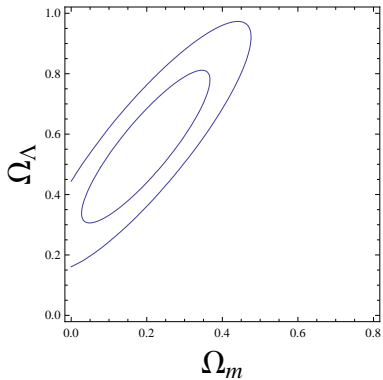}
\end{center}
\caption{ We plot the contour at 1$\sigma$  and 2$\sigma$ for the
parameters $\Omega_m$ and $\Omega_{\Lambda}$ after marginalization
of $h$. Here we have used the compressed data from
\citep{jla}.}\label{fig:figure00}
\end{figure}

\subsection{Intermediate model}

In order to compare with the results of the paper
by \cite{Dolgov:2014faa}, here we describe the results using both SNIa
data sets, together the BAO points.

\begin{figure}[h!]
\begin{center}
\includegraphics[width=5cm]{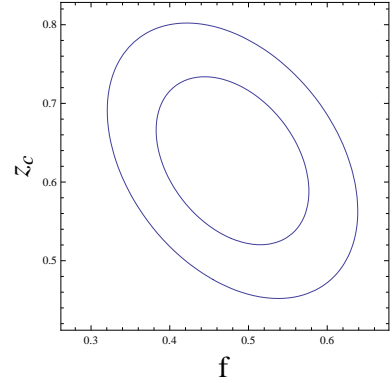}
\end{center}
\caption{ We plot the contour at 1$\sigma$  and 2$\sigma$ for the
parameters $f$ and $z_c$ after marginalization of $h$. Here we have
used data from the Loss compilation set
\citep{Ganeshalingam:2013mia} and the BAO points discussed in the
text.}\label{fig:figure01}
\end{figure}

The best fit values of the parameters -- using the Loss set together
with the BAO points -- are: $f=0.48\pm0.06$, $z_c=0.63\pm0.07$ and
$h=0.675\pm0.004$. Actually, these values remains essentially the
same with and without taking the three BAO points into consideration.
This fact is also mentioned by the authors of \citep{Dolgov:2014faa}
in the analysis they performed. After marginalization of the
parameter $h$, we plot the confidence limits of the parameters $f$
and $z_c$ in Fig.\ref{fig:figure01}.

We follow \citet{Dolgov:2014faa} and use the compressed form of the
JLA likelihood where only $M$ is taken as a free parameter that is
constrained together with the cosmological parameters. This
procedure is equivalent to what we already did in the previous
case using the Loss compiled data set. In that case, we consider
$h$ as a free parameter, and after the analysis was performed, it
was marginalized. This is the well-know degeneracy between the
absolute magnitude and the Hubble constant.

\begin{figure}[h!]
\begin{center}
\includegraphics[width=5cm]{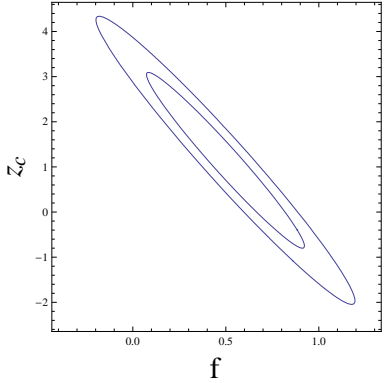}
\end{center}
\caption{ We plot the contour at 1$\sigma$  and 2$\sigma$ for the
parameters $f$ and $z_c$ after marginalization of $M$. Here we have
used data only from the JLA set \citep{jla}.}\label{fig:figure02}
\end{figure}

Using the JLA data set \citep{jla} alone, the best fit parameters
are: $f=0.50\pm0.28$, $z_c=1.2\pm1.3$ and $M=43.168\pm0.024$. Notice
the large uncertainty in the determination of the cross redshift
$z_c$. After marginalization of the nuisance parameter $M$, the
confidence contour of the parameters $f$ and $z_c$ are displayed in
Fig.(\ref{fig:figure02}).

\begin{figure}[h!]
\begin{center}
\includegraphics[width=5cm]{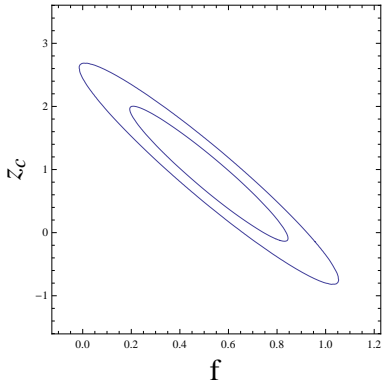}
\end{center}
\caption{ We plot the contour at 1$\sigma$  and 2$\sigma$ for the
parameters $f$ and $z_c$ after marginalization of $M$. Here we have
used SNIa data from the JLA \citep{jla} and the BAO
points.}\label{fig:figure03}
\end{figure}

After adding the BAO points, the result gives small changes compared
with the previous case with SNIa alone, with the following best-fit
parameters: $f=0.52\pm0.21$, $z_c=0.93\pm0.71$ and
$M=43.175\pm0.024$. After marginalization of the nuisance parameter
$M$ we get the confidence contours for the parameters $f$ and $z_c$
displayed in Fig.(\ref{fig:figure03}). It is clear that adding the
BAO points enables us to increase slightly the precision in the
determination of the free parameters.

A final comment would be useful here. A direct comparison between
Fig.(\ref{fig:figure01}) and Fig.(\ref{fig:figure03}) is not
completely fair. In fact, the LOSS sample we have used, does not
include the estimated systematic errors, whereas the JLA set does.
This is the reason for the difference between the precision in the
parameter determination observed in the graphs.

\section{Discussion}

From a theoretical point of view, we know that the intermediate
evolution is more appropriated to describe the low redshift
evolution, because this allows the existence of a transition from
the decelerated expansion phase to the current accelerated
expansion phase.

In previous sections, we showed that the intermediate evolution
successfully fit the data from SNIa and BAO. In fact, for the best
fit using SNIa (JLA set) only, we get $\chi^2_{min}=32.89$ for the
intermediate evolution (with three free parameters, $f$, $z_c$ and
$M$), whereas within the $\Lambda$CDM (with also three free
parameters ($\Omega_m$, $\Omega_{\Lambda}$ and $h$)) we get
$\chi^2_{min}=32.76$, both certainly a good fit to the data (for the
binned JLA set with 31 points, the estimated variance in $\chi^2$ is
$\sigma \simeq \sqrt{2/31}\simeq 0.25$, which means that both fit
are comparable to each other (see \citet{andrae})).

\begin{figure}[h!]
\begin{center}
\includegraphics[width=7cm]{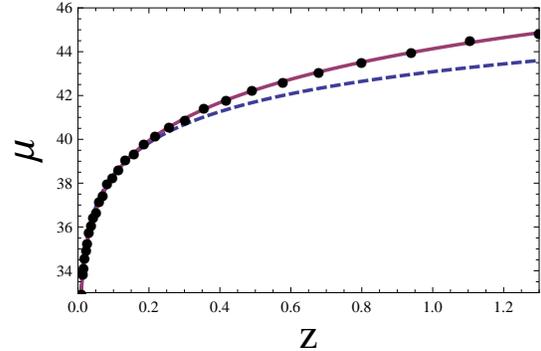}
\end{center}
\caption{ We plot the 31 data points from the JLA set \citep{jla}
together with the best-fit value found in this work, using the
intermediate evolution (\ref{a}) (continuous line), and the best fit
found in \citep{Dolgov:2014faa} using a power-law evolution $a(t)
\simeq t^{\beta}$ with $\beta=1.55$ quoted from that paper (dashed
line). Notice how the power-law fit moves away the data points from
$z>0.2$. }\label{fig:figure04}
\end{figure}

Just to make the comparison more evident, we plot the binned data
from the JLA set \citep{jla}, with the best fit curve obtained from
the theoretical model we studied -- the intermediate evolution
Eq.(\ref{a}) -- and the power-law model $a(t) \simeq t^{\beta}$ with
$\beta=1.55$ (quoted from \citep{Dolgov:2014faa}) in
Fig.(\ref{fig:figure04}).

It is clear that the intermediate evolution provides a better fit to
the data, than the power-law with $\beta=1.55$.

Actually, our study suggests that if we only use SNIa data, using all the data from the LOSS compiled set, and independently using all the data from the compressed JLA set, the power-law best fit does not correspond to $\beta \simeq 3/2$ rather we obtain $\beta = 1.78 \pm 0.09$ with LOSS, whereas $\beta = -0.35 \pm 0.06$ with JLA.

In this paper we have studied the intermediate evolution as the most 
appropriate for describing the low redshift regime supported by
observations from type Ia supernovae and BAO. We found that the
recent data (using the LOSS compiled sample
\citep{Ganeshalingam:2013mia} and JLA set \citep{jla}) suggest that
the intermediate evolution characterized by the scale factor
(\ref{a}) is as good a fit to this redshift regime as the standard
$\Lambda$CDM model.

This result can be considered a step forward from
\citep{Dolgov:2014faa}, where the authors used a power-law type
evolution $a(t) \simeq t^{\beta}$ and studied the constraints the
data imposed on the parameter $\beta$. Although they successfully
produce a good fit to the SNIa data, the model they consider is
unable to describe a transition from a decelerated to an accelerated
expansion phase. This transition is a key ingredient that any model
must satisfy if it claims to describe the recent (low redshift
$z<1$) evolution of the universe, which is essentially the same
redshift range as the supernova data span.

Although these results seems to rule out the power-law evolution to
describe the low redshift regime (unless we are interested in
fitting the very low regime $z<0.2$), the original motivation, that
of confronting modified gravity theories to data looking for signals
in favor of a dynamic adjustment mechanism to the vacuum energy
problem, remains intact. Actually, this becomes even more interesting
with the connection with the Gauss-Bonnet coupling terms needed in the
action to get the intermediate evolution.

\acknowledgments

VHC acknowledges financial support through DIUV 50/2013 and FONDECYT
1110230. This work is dedicated to the memory of our friend and
colleague Sergio del Campo with whom this project was initiated in
July 2014.

\end{document}